# Magnetocaloric effect in manganites: metamagnetic transitions for magnetic refrigeration

M. Quintero[1,2], J. Sacanell[1,*], L. Ghivelder[3], A. M. Gomes[3], A. G. Leyva[1,2] and F. Parisi[1,2]

[1] Gerencia de Investigación y Aplicaciones, Centro Atómico Constituyentes, CNEA, Av. Gral. Paz 1499, San Martín, Buenos Aires, Argentina.

[2] Escuela de Ciencia y Tecnología, Universidad Nacional de General San Martín, San Martín, Buenos Aires, Argentina.

[3] Instituto de Fisica, Universidade Federal do Rio de Janeiro, C.P. 68528, Rio de Janeiro, RJ, 21941-972, Brazil

We present a study of the magnetocaloric effect in $La_{5/8-y}Pr_yCa_{3/8}MnO_3$ (y=0.3) and $Pr_{0.5}Ca_{0.09}Sr_{0.41}MnO_3$ manganites. The low temperature state of both systems is the result of a competition between the antiferromagnetic and ferromagnetic phases. The samples display magnetocaloric effect evidenced in an adiabatic temperature change during a metamagnetic transition from an antiferromagnetic to a ferromagnetic phase . As additional features, $La_{5/8-y}Pr_yCa_{3/8}MnO_3$ exhibits phase separation characterized by the coexistence of antiferromagnetic and ferromagnetic phases and $Pr_{0.5}Ca_{0.09}Sr_{0.41}MnO_3$ displays inverse magnetocaloric effect in which temperature decreases while applying an external magnetic field. In both cases, a significant part of the magnetocaloric effect appears from non-reversible processes. As the traditional thermodynamic description of the effect usually deals with reversible transitions, we developed an alternative way to calculate the adiabatic temperature change in terms of the change of the relative ferromagnetic fraction induced by magnetic field. To evaluate our model, we performed direct measurement of the sample's adiabatic temperature change by means of a differential thermal analysis. An excellent agreement has been obtained between experimental and calculated data. These results show that metamagnetic transition in manganites play an important role in the study of magnetic refrigeration.

---

[*] Corresponding author: sacanell@cnea.gov.ar

The magnetocaloric effect (MCE) is characterized by the isothermal change of the magnetic entropy and the adiabatic change of temperature, induced by the application of an external magnetic field [1]. Due to the possibility of performing cycles of heat absorption and release just by varying a magnetic field on a magnetic material, the effect is considered as an alternative to develop more efficient and environmentally friendly cooling systems than current ones based on expansion and compression cycles of potentially harmful gases.

Among systems which exhibit MCE, the compounds known as manganites [2] are highly promising [3,4,5,6,7,8]. Due to the strong coupling between their magnetic, electronic and structural degrees of freedom, some manganites display the phenomenon of phase separation (PS), characterized by an intrinsic coexistence of ferromagnetic (FM) and antiferromagnetic (AFM) phases. Within the PS regime, the transition temperatures strongly depend on doping [9]. Since the MCE is maximum near $T_C$, it is straighfoward to obtain manganites for which the effect is enhanced at different temperatures. This property is unique because in a practical system, it can be used to broaden the refrigeration temperature range by developing composite systems formed by materials with similar structure, thus reducing the problems related to mechanical compatibility[10].

Metamagnetic transitions between AFM and FM states can be found within PS, during which a substancial amount of heat is released. Moreover, a substancial enhancement in MCE was recently observed in the PS regime [11,12]. As a counterpart, dynamical evoultion of the physical properties was also observed in some PS manganites, undobtedly suggesting its out of equilibrium nature [13,14]. This complicates the analysis of MCE in manganites with PS, typically based on the use of the thermodynamic Maxwell relations, which are derived for a homogeneous system in equilibrium. In that case, the use of the traditional approach to describe MCE through the entropy change [1] must be carefully revised. In fact, it has been recently shown that the use of Maxwell relations to analyze MCE in PS manganites, can either overestimate or underestimate their actual entropy change [6]. This problem was also observed in the $Mn_{1-x}Fe_xAs$ system, in which the inadequate use of the Maxwell relations for MCE calculation in metamagnetic transitions, leads to an overestimation of the entropy[15]. The authors propose a method that uses the Clausius-Clapeyron equation to calculate the MCE. However, no comparison with direct measurements is presented there.

In this work we present a study of the MCE in two manganites, $La_{5/8-y}Pr_yCa_{3/8}MnO_3$ with y=0.3 (LPCMO) and $Pr_{0.5}Ca_{0.09}Sr_{0.41}MnO_3$ (PCSMO). The first is a prototypical manganite exhibiting PS [9,16,17,18,19], while the second experiences transitions between FM and AFM states on cooling [20] and endothermic metamagnetic transitions, displaying inverse MCE (IMCE) [21]. Both systems undergo irreversible metamagnetic transitions in which the proportion of AFM and FM phases vary when applying a magnetic field. It is largely reported that these transitions have structural character [20,24] and involve the unblock of metastable states [14,22]. The traditional description of MCE does not consider specifically this typically nonreversible contribution. However, a study of these processes has to be made in order to fully understand MCE on manganites.

We have performed magnetization and specific heat measurements to calculate MCE, and a differential thermal analysis (DTA) for its direct measurement. The study was conducted in the temperature range in which both systems present AFM to FM metamagnetic transitions. We further developed a phenomenological model which accounts for the measured MCE of both systems.

The synthesis of our polycrystalline samples is described elsewere [20,23]. Magnetization, specific heat and DTA measurements were performed on a Quantum Design PPMS platform. A system consisting of two Pt resistances (Pt1000) mounted on the PPMS sample holder was adapted for DTA measurements, as schematically shown in Fig. 1. The compound under study and a reference sample (a piece of alumina) are in thermal contact with each of the Pt resistances. A teflon layer separates the PPMS puck and the thermometers to optimize thermal insulation of the sample and the reference. Measuring simultaneously both thermometers we obtain the local temperature of each sample as the magnetic field is applied. The differential analysis is performed to get rid of any magnetoresistance within the Pt sensor and small thermal fluctuations, common to both thermometers.

In Fig. 2(a) we show magnetization ($M$) and specific heat ($C_p$) data corresponding to the LPCMO sample. The system is a paramagnetic insulator at room temperature. On lowering $T$, an insulating charge ordered (CO) phase develops at $T_{co}$ ~ 220 K, as evidenced by peaks both in $M$ and $C_p$. On further cooling, nucleation of FM droplets occurs below $T_{C1}$ ~ 220 K, which coexists with the CO phase in the 220–80K temperature range [24]. Below $T_{C2}$ ~ 100 K, a long-range FM phase is formed [17]. No peak or discontinuity is observed at this temperature in the $C_p$ data, suggesting that the increase observed in $M(T)$ data does not correspond to a true phase transition. The PCSMO compound goes through two phase transitions on cooling (Fig. 2(b)), as evidenced by both *M* and $C_p$ measurements. At $T_C$ = 250 K, a FM phase develops. A CO-AFM phase appears below $T_{co}$ = 180 K.

In what follows we show the results of the study of MCE for both compounds when the CO-AFM to FM metamagnetic transition is induced by the application of an external magnetic field. We explore the behavior of *M* and the field induced temperature change *ΔT* (with respect to the target temperature) as a function of the applied field *H*. It is worth noting that these measurements were performed independently, because of the different experimental setups. Figure 3(a) shows the *M* vs *H* dependence at 110 K for the LPCMO sample, in which a crossing occurs from a phase separated state to an almost homogeneous ferromagnetic one as a function of H. Superimposed with the M vs H data, we show *ΔT* occuring during the metamagnetic transition. As a distinctive fact, the sample remains in its FM state after the applied field is completely removed.

Fig. 3b display the equivalent measurements for PCSMO at T= 160K. The M vs H curves present metamagnetic transitions with a saturation value close to that expected for a colinear FM sample. We observe an hysteretic behavior, with a field $H_M^+$ for the CO to FM transformation and a $H_M^-$ ($<H_M^+$) for the reverse transition. We also show *ΔT* for the metamagnetic transition, in which the IMCE can be seen both in H rise and fall. The values obtained for *ΔT* are within those observed for other manganites at similar magnetic field values[25,26].

To calculate the adiabatic temperature change induced by the application of the external magnetic field we estimated the change of the magnetic enthalpy instead of considering the entropy change. The main reason for this choice is that since we are dealing with a metamagnetic transition, we must find some link between the two pure phases through a thermodynamic state function. In the case of the enthalpy this link is obtained from magnetization measurements[27].

For a sample with a relative FM fraction $x$, the total enthalpy can be written as $E = x.E_{FM} + (1-x)E_{CO}$, with $E_{FM}$ ($E_{CO}$) the enthalpy of the pure FM (CO) phase. We assumed that the only effect of the magnetic field is on $E_{FM}$, as $E_{FM}(H) = E_{FM}(0) - M_S H$ ($M_S$: magnetization of an entirely FM sample). This fact is supported by the negligible influence that $H$ has on the $T_{co}$ of similar systems [9,27]. An increase in the magnetic field induces a growth of the FM relative fraction, and a release of heat within the sample, proportional to the difference of the enthalpies of the coexisting phases ($E_0$). Taking as a main approximation that this difference is independent of temperature, this process yields, for an isolated sample, to a temperature increase of $\Delta T_{ad} = -\frac{1}{C}\left[\frac{\partial x}{\partial H}(E_0 - M_S H) - xH\frac{\partial M_S}{\partial H}\right]\Delta H$, where $C$ is the specific heat. As our sample has a non-negligible thermal coupling with the environment, a dissipation term must be included in order to compare our model with experimental data. The balance between the released and the dissipated heat gives rise to a temperature change of $\Delta T = \frac{\Delta T_{ad}}{C_{eff}} - \gamma(T-T_0)\Delta H$, where $\gamma$ is a thermal coupling parameter between the sample and the environment and $C_{eff}$ is a correction factor to account for the effective heat capacity of the sample holder. We used $E_0 = -28.3$ J/mol for LPCMO which was taken from Ref. 27 and $E_0 = 300$ J/mol for PSCMO, obtained through the $H$ and $T$ dependence of $T_{CO}$. $\gamma$ was extracted from the thermal relaxation of the sample in measuring conditions. Figures 3(a) and 3(b) show the calculated value of $\Delta T$ for both samples in solid lines, displaying an excellent agreement with the measured data.

As recently reported [11,12], the MCE is enhanced on manganites with phase separation. The wide temperature range in which some manganites (as those presented here) display PS, and the facility to change this range by grain size [28,29], doping [30], etc, makes these systems particularly interesting for magnetic refrigeration. We showed in the present study that a significant part of the MCE in PS manganites arises from a nonreversible transition which is usually not considered in the analysis of the MCE phenomenon. Our model considers this contribution, allowing a complete description of the obtained data both for the direct and indirect measurements of the MCE. Additionally, it shows that $\frac{\partial x}{\partial H}$ and $\frac{\partial M_S}{\partial H}$ are key parameters to look for large MCE in PS manganites, both of which can be easily obtained from magnetization measurements. The DTA technique used has shown to be suitable to eliminate undesired experimental artifacts, typically originated by the thermometers used to measure the MCE. The present work opens a path for a thorough study of MCE in the metamagnetic transition of manganites exhibiting PS, in order to develop tunable materials for future magnetic refrigeration systems.

**Acknowledgments**

M. Quintero and J. Sacanell are members of CIC CONICET, both authors equally contributed to this work. Support from ANPCyT PICT 06-01549, and CONICET PIP-00038/08 is acknowledged. A.M Gomes and L. Ghivelder acknoledge financial support from CNPq and Faperj. The PCSMO sample was kindly given by F. Damay.

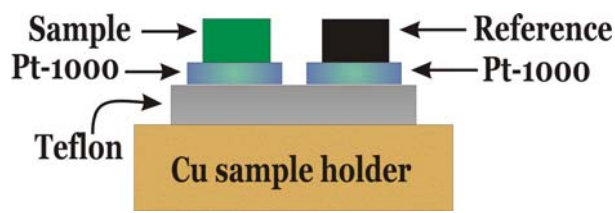

**Figure 1: Schematic picture of the system developed for the differential thermal analysis. The sample and the reference (a piece of alumina) are mounted on the Pt thermometers. The system is mounted on a Quantum Design PPMS, Cu sample holder.**

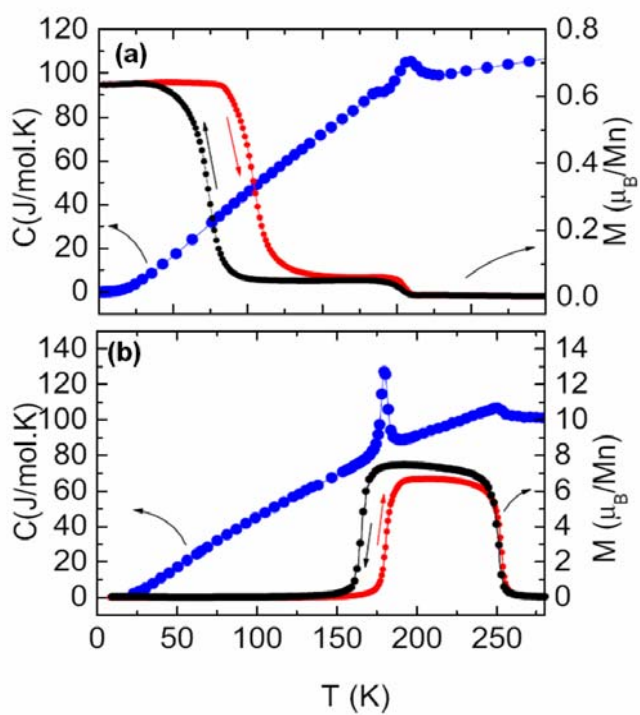

**Figure 2: Magnetization and Specific Heat as a function of temperature for (a) the LPCMO and (b) the PCSMO system.**

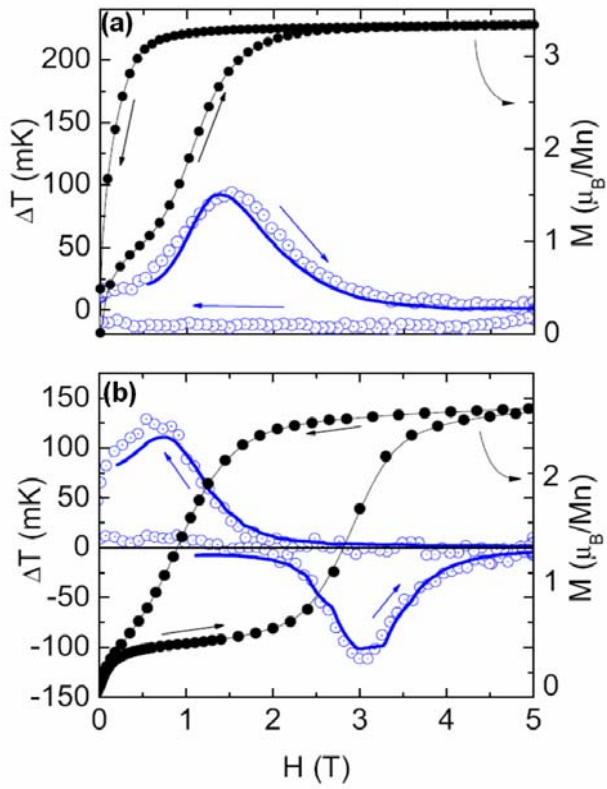

**Figure 3: Magnetization and ΔT vs H for (a) LPCMO at 110 K and (b) PCSMO at 160 K. Filled symbols corresponds to magnetization. Open symbols indicates the measured ΔT and lines show the results extracted from the model.**